\documentclass[aip,graphicx]{revtex4-1}
\usepackage[utf8]{inputenc}
\usepackage{amsmath}
\usepackage{amsfonts}
\usepackage{amssymb}
\usepackage{color}
\usepackage{graphicx}
\usepackage{hyperref}
\usepackage{lastpage}
\usepackage{fancyhdr}
\newcommand{\nn}{\nonumber}
\begin{document}
\title{Magnetic Monopoles in Noncommutative Quantum Mechanics 2}

\author{Samuel Kov\'a\v{c}ik}

\address{Dublin Institute for Advanced Studies, 10 Burlington Road, Dublin 4, Ireland}

\email[]{skovacik@stp.dias.ie}

\author{Peter Prešnajder}

\address{Faculty of Mathematics, Physics and Informatics, Comenius University Bratislava, Mlynsk\'a dolina, Bratislava, 842 48, Slovakia}

\email[]{presnajder@fmph.uniba.sk}

\date{\today}

\begin{abstract}
In this paper we extend the analysis of magnetic monopoles in quantum mechanics in three dimensional rotationally invariant noncommutative space $\textbf{R}^3_\lambda$. We construct the model step-by-step and observe that physical objects known from previous studies appear in a very natural way. Nonassociativity became a topic of great interest lately, often in connection with magnetic monopoles. We show that this model does not possess this property. 
\end{abstract}

\maketitle

\section{Introduction}  \label{sec1}
Despite never being actually observed, magnetic monopoles keep appearing in physical theories for over a century. Maxwell equations, derived to describe the classical electromagnetism, offer a straightforward generalization (or symmetrization if one prefers) which introduces magnetic monopoles. 

This newly promoted symmetry between the electric and the magnetic content of the theory seems to be lost in the formalism of electromagnetic potentials, where magnetic monopoles fields can be described by (line) singular potentials \cite{dirac}. It should be of no surprise as the formalism was constructed to have $\mbox{div } \textbf{B} = 0$ hardwired and only the singular behavior allows to avoid this (making the derivatives not commuting allows for $\mbox{div rot } \textbf{A} = 4 \pi \rho_M \neq 0$). 

This behavior persists in the quantum description. As was shown by Zwanziger \cite{zwanziger}, a system containing a magnetic monopole is described by a deformed Heisenberg algebra with non-vanishing commutator of conjugate momenta. Study of such systems is alluring because of many novel features, for example a static system of electric and magnetic charge generates a field with nontrivial angular momentum. 

As was shown by Polyakov and 't Hooft \cite{pol, hooft}, the existence of magnetic monopoles is a rather general consequence of the Grand Unified Theory as they are being formed when the higher symmetry breaks down into a product containing $U(1)$. They also appear as topological solutions in SYM theories \cite{Seiberg:1994rs} and in M-theory, where they can be lifted into higher dimensions \cite{Gauntlett:1992nn}, or as certain gravitational solutions in Kaluza-Klein theories \cite{Sorkin:1983ns,Gross:1983hb} . 

There is one, even though just a rather indirect, evidence of magnetic monopoles. A product of electric and magnetic charge has to satisfy the Dirac quantization condition so the existence of magnetic monopoles would imply that electric charge has to be quantized as well – as is indeed observed in nature. 

A plausible overall picture is the following: magnetic monopoles do exist, but are too heavy (on the GUT scale) to be produced in particle colliders. Those created shortly after the Big bang were diluted by the process of inflation, yet are still present and therefore explain discreteness of the electric charge. Of course, another explanation is viable as well – they just might not exist at all. However, we will follow the optimism of Polchinski \cite{Polchinski:2003bq} and many others – assume magnetic monopoles do exist and investigate them in the context of quantum mechanics in noncommutative space. 

Noncommutative (NC) space is a space whose close points cannot be distinguished. The name comes from the nonzero commutator of the coordinate operators, an analogical situation to the ordinary quantum mechanics (QM) with noncommutative phase-space. 

Theories in NC spaces were originally considered as a way of controlling UV divergences as by restricting infinitely short distances one also eliminates infinitely large energies \cite{groenewold,snyder, fuzzy}. This task was later taken over by the program of renormalization and the interest in NC theories diminished for many years and has been revitalized only rather recently by Connes and others \cite{Con}. 

Nowadays, NC spaces are an important feature of different theories of quantum gravity and emergent space(time) where they often pose as a middle-point between an ordinary space and more fundamental objects. These approaches often predict a nontrivial space structure below the Planck scale, but there are also examples where NC works as an effective description of the underlying physics on more ordinary scales \cite{Fujii:2005kg,hall}. 

We study consequences of the space noncommutativity in the context of QM. Our goal is to examine to what extent does the formalism have to be adapted (recall the great shift between the ordinary mechanics and the QM originating from the phase space noncommutativity), whether the theory remains self-consistent and offers some new prospects and properties. 

In our previous works we have shown that the hydrogen atom problem remains exactly solvable as contrary to lattice discretizations the relevant symmetries remain unspoilt \cite{GP1,GP2}, and that the expected UV regularization takes an explicit form which points towards higher structures present in the theory \cite{vel}. We have also noticed that magnetic monopoles appear as a very natural generalization of the considered Hilbert space of states  \cite{mm}. In this paper we will investigate this issue in more detail. 

The model of NC space used in this paper can be understood as a sequence of concentric fuzzy spheres of increasing radius. Magnetic monopoles on a single fuzzy sphere has been studied for example in \cite{mms1, mms2} and references therein. Dirac quantization condition in NC space-time has been analyzed in \cite{mmDQC}.

The paper is organized as follows: in section \ref{sec2} we repeat the construction of $R_\lambda^3$, in section \ref{sec3} we construct an algebra of operators consistent with the monopole structure. In section \ref{sec4} we investigate the velocity operator and its dual. The last section \ref{sec5} is devoted to conclusions, after which the appendix follows.

\section{Quantum mechanics in noncommutative space $ \textbf{R}^3_\lambda$} \label{sec2}
To build a NC space we need a NC tool. There are many choices, for example an algebra of functions equipped with a NC product  \cite{groenewold,moyal} or an algebra of matrices \cite{fuzzy}. Our choice is the auxiliary operator construction from \cite{Jabbari}, which was developed in more detail in \cite{GP1,GP2, vel, mm,LRL,plane,mbh}. We will use two sets of creation and annihilation (c/a) bosonic operators satisfying the usual relations
\begin{equation}\label{aux}
[a_\alpha,a^+_\beta]=\delta_{\alpha\beta },\ \
[a_\alpha,a_\beta]=[a^+_\alpha, a^+_\beta]=0,
\end{equation}
with $\alpha, \beta =1,2$ and acting in an auxiliary space $\mathcal{F}$ spanned by normalizes states
\begin{equation}
|n_1,n_2\rangle= \frac{(a^+_1)^{n_1}\,(a^+_2)^{n_2}}{
\sqrt{n_1!\,n_2!}}\ |0\rangle.
\end{equation}
The Fock space is a sum $\mathcal{F} = \sum \limits_{n=0}^\infty \bigoplus \mathcal{F}_n $ where $\mathcal{F}_n$ contain states $|n_1,n_2\rangle$ with $n_1+n_2=n$. 

The simplest nontrivial operators $\mathcal{F}_n\rightarrow \mathcal{F}_n$ are those of the form $a^+_\alpha a_\beta$. We can contract their indices either using the Pauli matrices $\sigma^i_{\alpha \beta}$ or the Kronecker symbol $\delta_{\alpha \beta}$ to obtain (after minor modifications)
\begin{equation} \label{X}
x_i = \lambda \sigma^i_{\alpha \beta} a^+_\alpha a_\beta, r = \lambda (a^+_\alpha a_\beta +1).
\end{equation}
$\lambda$ has been added to introduce length scale and $+1$ was added to ensure that $x^2 - r^2 = O(\lambda^2)$. These are the coordinates of the noncommutative (NC) space $\textbf{R}_\lambda^3$, $x^i$ taking the role of Cartesian coordinates and $r$ being the radial distance from the origin. They satisfy the following relations
\begin{equation} \label{NCrel}
[x_i, x_j] = 2 i \lambda \varepsilon_{ijk} x_k,\ [x_i, r] = 0, \ x^2 = r^2 - \lambda^2.
\end{equation}
$\lambda$ is the constant of noncommutativity, describing the length scale under which one cannot distinguish two close points of space. In physical applications it is assumed to be approximately the Planck length $\lambda \sim l_P \approx 1.6 \times 10^{-35}m$. 

To study quantum mechanics in $\textbf{R}_\lambda^3$ we will consider a Hilbert space of states $\mathcal{H}_\kappa$ consisting of functions of the form $\Psi_\kappa(a,a^+)$ satisfying 
\begin{equation} \label{states}
\Psi_\kappa(e^{-i\tau} a^+, e^{i\tau} a) = e^{-i\tau\kappa} \Psi_\kappa (a^+,a), \ \tau \in  \textbf{R}, \  \mbox{fixed}\  \kappa \in \textbf{Z}  ,
\end{equation}
and equipped with the scalar product 
\begin{equation} \label{norm}
(\Phi_\kappa,\,\Psi_\kappa)\ =\ 4\pi \lambda ^2 Tr [ \Phi^+_\kappa \,\hat{r}\, \Psi_\kappa] , \ \ \ \hat{r} \Psi = \frac{1}{2}\left( r \Psi + \Psi r \right)\, ,
\end{equation}
$\kappa$ is an integer, the difference in the number of creation and annihilation operators in $\Psi_\kappa (a,a^+)$. Hermitian conjugated operator $\hat{\mathcal{O}}^\dagger$ with respect to the scalar product \eqref{norm} is defined as usual
\begin{equation} \label{conj}
(\Phi_\kappa,\,\hat{\mathcal{O}} \Psi_\kappa)\ =\ ( \hat{\mathcal{O}}^\dagger \Phi_\kappa,\,\Psi_\kappa)\, .
\end{equation}

Note that the NC coordinates \eqref{X} contain an equal number of creation as annihilation operators. Therefore, the subspace $\mathcal{H}_{\kappa=0}$ contains states that have commutative counterparts of the form $\psi(x)$. The rest, $\mathcal{H}_{\kappa \neq 0}$, contains monopole states of field strength $\mu = -\frac{\kappa}{2}$, see \cite{mm}. Note that this relation makes the Dirac quantization condition $\mu \in \textbf{Z}/2$ satisfied in a very natural way. 

There is a different point of view on this construction of NC QM which provides a deeper insight and also a commutative counterpart of the theory \cite{comMM}. 

The starting point is to realize that the three-dimensional (commutative) Euclidean space $\textbf{R}^3$ is closely related to the complex dimensional space $\textbf{C}^2$. Even thought the number of dimensions differs, their symmetry groups (of rotations) are locally isomorphic. Two complex coordinates $z_\alpha$ of $\textbf{C}^2$ can be mapped into three real coordinates $x_i$ of $\textbf{R}^3$ as 
\begin{equation} \label{hopf}
x_i = \bar{z} \sigma_i z,
\end{equation}
with $\sigma^i$ being the usual Pauli matrices. This is a complex Hopf fibration, as can be seen by using Cayley-Klein parameters to describe $S^3$ spheres in $\textbf{C}^2$ being mapped into $S^2$ spheres in $\textbf{R}^3$. 

$\textbf{C}^2$ is naturally equipped with a Poisson structure $\{ z_\alpha, z^+_\beta \}_P = -i\delta_{\alpha \beta}$ which allows a straightforward quantization. To do so one has to replace the (complex) coordinates with c/a operators acting in an auxiliary Fock space as $z_\alpha \rightarrow \sqrt{\lambda} a_\alpha, \bar{z}_\alpha \rightarrow \sqrt{\lambda} a_\alpha^+$, where $\lambda$ has the dimension of length and to replace Poisson brackets with commutators $\{ \cdot \ , \ \cdot \}_P \rightarrow -i [ \cdot \ , \ \cdot ]$. The relation $x_i = \bar{z} \sigma_i z$ carries this quantization into $\textbf{R}^3$, creating $\textbf{R}^3_\lambda$.

Everything we do in NC QM has a commutative counterpart that can be obtained by going the other way – replacing c/a operators with complex coordinates and commutators with Poisson brackets. 

One can formulate QM in $\textbf{C}^2$, the free Hamiltonian $\hat{h}_0$ and the velocity/momentum operator $\hat{v}_i$ can be constructed using the Poisson structure, see \cite{mm}. If one defines the states as $L^2$ functions of the form $\psi(\textbf{x})$ with $x_i$ defined in \eqref{hopf}, actions of the Hamiltonian and the momentum operator mimic their $\textbf{R}^3$ counterparts, for example: $\hat{h}_0 \psi(x) \propto \partial_i \partial_i \psi(x), \hat{v}_i \psi(x)\propto i \partial_i  \psi(x)$. By restricting only on the functions of these specific combination of $\bar{z}, z$, one recreates the ordinary $\textbf{R}^3$ QM, but in $\textbf{C}^2$. 

However, if after doing so a more general class of states is considered, with $\Psi_\kappa = \psi(\textbf{x}) z_1^{\kappa_1} z_2^{\kappa_2}|_{-\kappa = \kappa_1 + \kappa_2}$, $\textbf{R}^3$ QM with magnetic monopoles of strength $\mu = - \kappa / 2$ is realized. A good example of this is that $[\hat{v}_i, \hat{v}_j] \Psi_\kappa = -i\frac{\kappa}{2} \frac{\varepsilon_{ijk} \hat{x}_i}{r^3} \Psi_\kappa$, where $r^2 = x_ix_i$, which is to be compared with the result of Zwanziger \cite{zwanziger} for the commutator of conjugate momenta of a system with magnetic monopole $[\hat{\pi}_i, \hat{\pi}_j] = i \mu \frac{\varepsilon_{ijk}  \hat{x}_k}{r^3}$.

How is it possible that generalized states of QM in $\textbf{C}^2$ describes monopole states in QM in $\textbf{R}^3$? The answer is that the angular coordinate $\gamma$, which is lost in the Hopf fibration \eqref{hopf}, persists in $\kappa$ states as $\Psi_\kappa = \psi(x) e^{- i \frac{\kappa}{2} \gamma}$ and serves as an extra compact direction the solution can wind around (an integer amount of times). 

This concludes our discussion of the commutative counterpart of the theory, we shall now return to the center of our interest – QM in $\textbf{R}_\lambda^3$.

\section{Quadratic operators on $\mathcal{H}_\kappa$}  \label{sec3}

Defining the Hilbert space $\mathcal{H}_\kappa$ equipped with a norm is just the first part of constructing (NC) QM. The other is to introduce operators which provide the physical meaning of the theory. 

As both the underlying NC space $\textbf{R}_\lambda^3$ and the Hilbert space $\mathcal{H}_\kappa$ are realized using c/a operators \eqref{aux} it shall be of no surprise that the same auxiliary operators can be used to define the operators on $\mathcal{H}_\kappa$.

The simplest possible action would be to take just one auxiliary operator and add it either on the left or the right side of $\Psi_\kappa$:
\begin{eqnarray} \label{aux2}
\hat{a}_\alpha \Psi_\kappa = a_\alpha \Psi_\kappa,  &&\ \ \ \hat{a}_\alpha^+ \Psi_\kappa = a_\alpha ^+ \Psi_\kappa , \\ \nn
\hat{b}_\alpha \Psi_\kappa =  \Psi_\kappa a_\alpha, &&\ \ \ \hat{b}_\alpha^+ \Psi_\kappa =  \Psi_\kappa a_\alpha ^+ .
\end{eqnarray}
Note that $[\hat{a}_\alpha, \hat{a}_\beta^+ ] = - [\hat{b}_\alpha , \hat{b}_\beta^+] = \delta_{\alpha \beta}$. It should be stressed that due to the factor $\hat{r}$ in the norm (\ref{norm}) the operators $\hat{a}_\alpha^+,\ \hat{b}_\alpha^+$ are not Hermitian conjugated to $\hat{a}_\alpha,\ \hat{b}_\alpha$ respectively. 

As $\kappa$ denotes the difference in the number of creation and annihilation operators in $\Psi_\kappa$, actions defined in \eqref{aux2} maps $\mathcal{H}_\kappa \rightarrow \mathcal{H}_{\kappa \pm 1}$. To stay in the same Hilbert subspace a creation operator has to always be paired with an annihilation one (and vice versa). Therefore, the simplest operators are quadratic actions of \eqref{aux2}, possibly with a factor of $\hat{r}$ to achieve Hermiticity under \eqref{norm}.

In hindsight, most of the operators investigated in the previous studies of QM in $\textbf{R}_\lambda^3$ are of this form. For example the angular momentum operator in \cite{GP1, GP2}, the position and the velocity operator in \cite{vel} or the Laplace-Runge-Lenz vector in \cite{LRL}. It has been also noted that these operators often form interesting algebraic structures, for example $so(1,3)$ and $so(4)$ of the angular momentum and the Laplace-Runge-Lenz vector in \cite{LRL} or $so(4)$ of the angular momentum operator and the coordinate operators in \cite{vel}. We will shortly reveal that these all were just parts of a larger scheme.

To do so we will (re)construct these operators in a new way, making the underlying symmetries completely transparent. Let us first define a set of $ 4 \times 4 $ matrices $S_{AB}=-S_{BA}$ satisfying the $su(2,2)$ algebraic relations. To avoid confusion, the range of indices is: $A,B, ... = 0, ..., 5$; $a,b, ... = 1, ..., 4$ and $i,j, ... = 1,2,3$.
\begin{eqnarray}
S_{ij} = \frac{1}{2}\varepsilon_{ijk} \left( \begin{array}{cc}
\sigma_k & 0 \\ 
0 & \sigma_k 
\end{array} \right), && S_{k4} = \frac{1}{2} \left( \begin{array}{cc}
\sigma_k & 0 \\ 
0 & -\sigma_k 
\end{array} \right), \\ \nn
S_{0k} = \frac{i}{2} \left( \begin{array}{cc}
0& \sigma_k  \\ 
\sigma_k & 0
\end{array} \right), && S_{45} = \frac{i}{2} \left( \begin{array}{cc}
0 & 1 \\ 
1 &0 
\end{array} \right) , \\ \nn
S_{k5} = \frac{1}{2} \left( \begin{array}{cc}
0& \sigma_k  \\ 
-\sigma_k & 0
\end{array} \right), && S_{04} = \frac{1}{2}\left( \begin{array}{cc}
0 & 1 \\ 
-1 &0 
\end{array} \right) , \nn
\end{eqnarray}
\begin{equation}
S_{05} = \frac{1}{2}\left( \begin{array}{cc}
 1 & 0 \\ 
0& -1 
\end{array} \right) .
\end{equation}
These matrices satisfy the $su(2,2)$ relations (we are using the $so(4,2)$ notation with $\eta = \mbox{diag}(1,1,1,1,-1,-1)$, as it is isomorphic): 
\begin{equation} \label{Gamma}
 S^+_{AB} \ =\ - \Gamma\, S_{AB}\, \Gamma , \ \ \ \Gamma\ =\ \left( \begin{array}{cc}
 1 & 0 \\ 
0& -1 
\end{array} \right), 
\end{equation}
\begin{equation} [S_{AB}, S_{CD}]\ =\ i ( \eta_{AC} S_{BD} - \eta_{BC} S_{AD} - \eta_{AD} S_{BC} + \eta_{BD} S_{AC} ).
\end{equation}
Rotations are generated by 7 Hermitian matrices $S_{05}, S_{ab}$, while 8 anti-Hermitian matrices $S_{a5}, S_{0a}$ generate boosts.

We began with two c/a operators satisfying $[a_\alpha , a_\beta^+] = \delta_{\alpha \beta}$. Now, since these can be applied on $\Psi_\kappa$ either from the left or the right, see \eqref{aux2}, we effectively have four auxiliary c/a operators 
\begin{equation}
\hat{A}^T = (\hat{a}_1, \hat{a}_2, \hat{b}_1, \hat{b}_2) \,, \ \hat{A}^+ = (\hat{a}_1^+, \hat{a}_2^+, \hat{b}_1^+, \hat{b}_2^+) \,.
\end{equation}
Since the right multiplication exchanges the order, the commutator of $[b_\alpha , b_\beta^+]$ has an opposite sign compared to $[a_\alpha , a_\beta^+]$. This can be fixed using the $\Gamma$ matrix
\begin{equation}
[\hat{A}_a,  \Gamma_{bc}\hat{A}^+_c] =\delta_{ab},
\end{equation}
where $\Gamma$ is given in \eqref{Gamma}. Now, we can use $\hat{A}$ and $\Gamma \hat{A}^+$ to realize $su(2,2)$ operator representation using the matrix representation $S_{AB}$ as
\begin{equation} 
\hat{S}_{AB}\ =\ \hat{A}^+\, \Gamma\, S_{AB}\, \hat{A},
\end{equation}
or explicitly:
\begin{eqnarray} \label{S}
\hat{S}_{ij} &=& \frac{1}{2}\,\varepsilon_{ijk}\,(\hat{a}^+\,\sigma_k\,\hat{a}\,-\, \hat{b}^+\,\sigma_k\,\hat{b}) ,\ \ \  \hat{S}_{k4}\ =\ \frac{1}{2}\,(\hat{a}^+\,\sigma_k\,\hat{a}\,+\, \hat{b}^+\,\sigma_k\,\hat{b})\,, \nn \\ 
\hat{S}_{05} &=& \frac{1}{2}\,(\hat{a}^+\,\hat{a}\,+\, \hat{b}^+\,\hat{b}) ,\hskip2cm \hat{C}\ =\ \hat{a}^+\,\hat{a}\,-\, \hat{b}^+\,\hat{b}\,, \\ \nn
\hat{S}_{0k} &=& \frac{i}{2}\,(\hat{a}^+\,\sigma_k\,\hat{b}\,-\, \hat{b}^+\,\sigma_k\,\hat{a})\,, \ \ \ \hat{S}_{45}\ =\ \frac{i}{2}\,(\hat{a}^+\,\hat{b}\,-\, \hat{b}^+\,\hat{a})\,,\\ 
\hat{S}_{k5} &=& \frac{i}{2}\,(\hat{a}^+\,\sigma_k\,\hat{b}\,+\, \hat{b}^+\,\sigma_k\,\hat{a})\,, \ \ \ \hat{S}_{04}\ =\ \frac{1}{2}\,(\hat{a}^+\,\hat{b}\,+\, \hat{b}^+\,\hat{a})\,.
\end{eqnarray}
The rotations operators $\hat{S}_{ab}$ and $\hat{S}_{05}$ are combinations of operators $\hat{a}^+_\alpha  \hat{a}_\beta $ and $\hat{b}^+_\alpha \hat{b}_\beta $ that commute with $\hat{r}\,=\,\frac{\lambda}{2} (\hat{a}^+\hat{a}\,+\,\hat{b}^+\hat{b})$. We added the (Hermitian) central operator $\hat{C} = \hat{A}^+ \Gamma \hat{A}$, that specifies the Hilbert space in questions: $(\hat{C}+2) \psi_\kappa\,=\,\kappa\, \psi_\kappa$. Consequently, we obtained the unitary $u(2)\,\oplus\,u(2)$ Lie algebra representation in $\mathcal{H}_\kappa$.\\ 

The boosts operators $\hat{S}_{0a}$ and $\hat{S}_{a5}$ are linear combinations of $\hat{a}^+ \hat{b}$ and $\hat{b}^+ \hat{a}$ and therefore do not commute with $\hat{r}$. Consequently, they are non-Hermitian in $\mathcal{H}_\kappa$ with respect to \eqref{norm}. 


As has been mentioned already, the quadratic operators realizing $su(2,2)$ representation are closely related to physical operators known from previous studies of this model of NC QM. $\varepsilon_{ijk}\hat{S}_{jk}$ is proportional to the angular momentum operator $\hat{L}_i \sim \lambda^{-1} [ x_i , \ \cdot \ ]$, while $\hat{S}_{k4}$ is proportional to the coordinate operator $\lambda^{-1} \hat{X}_i$ (the need of symmetrized coordinate operator was inferred in \cite{vel} but also appears in a very different context in \cite{Deriglazov:2016mhk,Gomes:2010xk}). These operators are Hermitian under \eqref{norm} and require nothing else. 

The operators in the two bottom lines of \eqref{S} combine left and right actions and are not Hermitian under \eqref{norm}, which can be corrected by adding the factor of $r^{-1}$ to cancel the one in the norm. After doing so they can be related to the free Hamiltonian $\hat{r}^{-1} \hat{S}_{04} \sim \hat{V}_4\sim\hat{H}_0 +(...)$, the velocity operators $\hat{r}^{-1} \hat{S}_{0k} \sim \hat{V}_k$ and the Laplace-Runge-Lenz vector $\hat{r}^{-1} \hat{S}_{k5} \sim \hat{A}_k - (...)$. Actually, $\hat{r}^{-1} \hat{S}_{45}$ is the only not appearing in a physical context before – it is related to the dilation operator $\hat{D}=\hat{X}_i \hat{V}_i$. 

It shall be noted that adding the factor of $\hat{r}^{-1}$ has a drastic impact on the algebraic structure of the operators. For example $[\hat{S}_{0i}, \hat{S}_{0j} ] \sim \hat{S}_{ij}$ changes to $[\hat{r}^{-1} \hat{S}_{0i}, \hat{r}^{-1} \hat{S}_{0j} ] \sim \varepsilon_{ijk} \hat{S}_{4k} (\hat{C}-2)$, which is the relation that revealed the monopole behavior of $\Psi_\kappa$ states in \cite{mm}. 

Before moving forward, let us investigate the effect of including (functions of) $\hat{r}$ on the commutators. To do so we first define the following notation
\begin{equation}
\hat{\zeta}_a =  2\left( \hat{S}_{k5}, \hat{S}_{04}\right), \ \hat{w}_a = 2i\left(\hat{S}_{k 0},  \hat{S}_{54} \right), \delta_{\pm} f(\hat{r}) = f ( \hat{r} \pm \lambda).
\end{equation}
Using those it can be shown that 
\begin{eqnarray}
[\hat{w}_a, f(\hat{r})] &=&  \left(\frac{\Delta_+ + \Delta_- -2}{\lambda^2}f(\hat{r})\right) \frac{\lambda^2}{2}\hat{w}_a+ \left(\frac{\Delta_+ - \Delta_-}{2 \lambda}  f(\hat{r})\right)\lambda \hat{\zeta}_a \\ \nn
[\hat{ \zeta}_a, f(\hat{r})] &=&  \left(\frac{\Delta_+ + \Delta_- -2}{\lambda^2}f(\hat{r})\right)\frac{\lambda^2}{2}\hat{ \zeta}_a + \left(\frac{\Delta_+ - \Delta_-}{2 \lambda} f(\hat{r}) \right) \lambda \hat{w}_a
\end{eqnarray}
Note that in the commutative limit $\frac{\Delta_+ + \Delta_- -2}{\lambda^2}f(\hat{r}) \rightarrow d^2 f(r) / dr^2$ and $\frac{\Delta_+ - \Delta_-}{2 \lambda}  f(\hat{r}) \rightarrow d f(r) / dr$. However, we have to be careful since before taking the limit $d^2 f(r) / dr^2 \neq d (d f(r) / dr) / dr$ (they have different $\lambda$ shifts).

Let us now demonstrate the effect of $f(\hat{r})$ on the algebraic relations using the aforementioned example
\begin{equation} \label{[V,V]}
\varepsilon_{ijk}  [f(\hat{r})\hat{w}_i, f(\hat{r}) \hat{w}_j] = f(\hat{r}) \left(\mathcal{D}f(\hat{r})\right) 4 i \varepsilon_{ijk} S_{ij} + f(\hat{r})\left(\frac{\Delta_+ - \Delta_-}{2 \lambda}  f(\hat{r})\right) 4 i \lambda \hat{S}_{4k} \left(\hat{C}+2\right),
\end{equation}
where $\mathcal{D} =  \left(1 + \frac{\Delta_+ + \Delta_- -2}{2} +\hat{r} \frac{\Delta_+ - \Delta_-}{2 \lambda}   \right) $. It can be checked that $\mathcal{D}  \hat{r}^{-1}=0$, so the choice of $f(\hat{r}) = \hat{r}^{-1}$ cancels these terms and simplifies the relations considerably.

To conclude, the factors of $f(\hat{r})$ change the considered algebra significantly, the resulting algebra encloses, but requires an infinite tower of factors $\Delta_\pm ^n f(\hat{r})$. The choice of $f(\hat{r})= \hat{r}^{-1}$ not only ensures Hermiticity under \eqref{norm}, but also simplifies the algebra considerably. Let us assume it from now on.

\section{Velocity operator and its dual}  \label{sec4}
The velocity and the dual velocity operators are defined as
\begin{equation}
\hat{V}_a\ =\ 2\hat{r}^{-1}\,\hat{S}_{0a}\,, \ \ \ \tilde{V}_a\ =\ 2 \hat{r}^{-1}\,\hat{S}_{a5}\ =\ 2 i \hat{r}^{-1}[\hat{S}_{05}, \hat{S}_{0a}]\,, 
\end{equation}
or explicitly:
\begin{eqnarray}
\hat{V}_k\ =\ \frac{i}{2\hat{r}}\,\left( \hat{a}^+ \sigma_k \hat{b}\,-\,\hat{b}^+ \sigma_k \hat{a} \right) \,,\ \ \ \hat{V}_4\ =\ \frac{1}{2\hat{r}}\,\left( \hat{a}^+ \hat{b}\,+\,\hat{b}^+ \hat{a} \right)\,, \\ \nn
\tilde{V}_k\ =\ \frac{1}{2\hat{r}}\,\left( \hat{a}^+ \sigma_k \hat{b}\,+\,\hat{b}^+ \sigma_k \hat{a} \right) \,,\ \ \ \tilde{V}_4\ =\ \frac{i}{2\hat{r}}\,\left( \hat{a}^+ \hat{b}\,-\,\hat{b}^+ \hat{a} \right)\,.
\end{eqnarray}
The rotation $e^{i\omega\hat{S}_{05}}$ transforms $\hat{V}_a$ to $\hat{V}_a(\omega )\,=\,\cos\omega \,\hat{V}_a\,+\,\sin\omega \,\tilde{V}_a$. They can be expressed as linear combinations of the operators 
\begin{equation}  \label{U}
\hat{U}_{\alpha \beta}\ =\ \frac{1}{\hat{r}}\hat{a}^+_\alpha \hat{b}_\beta \,,\ \ \   \hat{U}^\dagger_{\alpha \beta}\ =\ \frac{1}{\hat{r}} \hat{a}_\alpha \hat{b}^+_\beta\, ,
\end{equation}
whose properties are studied in more detail in the appendix A1. The important result is that the only non-vanishing commutator is
\begin{equation} \label{UU*4}
 [\hat{U}_{\alpha \beta}, \hat{U}^\dagger_{\gamma \delta} ]\ =\ -\frac{1}{\hat{r}^2} \left( \hat{a}^+_\alpha \hat{a}_\gamma \,\delta_{\beta \delta} \ +\ \hat{b}_\delta^+\hat{b}_\beta \,\delta_{\gamma \alpha} \right) +\ \frac{\lambda}{\hat{r}}\,\{ \hat{U}_{\alpha \beta}, \hat{U}^\dagger_{\gamma \delta}\},  
\end{equation}
%
where $\{ \hat{A}\,,\hat{B} \}\,=\,\hat{A}\,\hat{B}\,+\,\hat{B}\,\hat{A}$. The $\hat{r}^{-1}$ factor surprisingly appears without the $\lambda$-shift. The resulting formula can be
rewritten as a $q$-deformed commutator with $\hat{r}$-dependent deformation parameter $\hat{Q} \,=\,(\hat{r}-\lambda)/(\hat{r}+\lambda)$, which approaches $1$ for $r\gg \lambda$:
\begin{equation} \label{UU*order}
\hat{U}^\dagger_{\gamma \delta}\,\hat{U}_{\alpha \beta}\ =\ \frac{\hat{r}-\lambda}{\hat{r}+\lambda }\,\hat{U}_{\alpha \beta}\, \hat{U}^\dagger_{\gamma \delta}\ -\ \frac{1}{\hat{r}^{2}(\hat{r}+\lambda)}\left( \hat{a}^+_\alpha \hat{a}_\gamma \,\delta_{\beta \delta} \,+\,\hat{b}_\delta^+\hat{b}_\beta \,\delta_{\gamma \alpha} \right) \,.   
\end{equation} 
%
Thus we have an associative complex algebra $\mathcal{U}$ generated by polynomials in $so(4)$ generators $\hat{S}_{ab}$ and operators $\hat{U}_{\alpha \beta}$, $\hat{U}^\dagger_{\alpha \beta}$  and analytic in $\hat{r}\,=\,\lambda^{-1} \hat{S}_{05}$. The defining relations of $\mathcal{U}$ are:\\

(i) $\hat{S}_{ab}$, $a,b\,=\,1,2,3,4$, satisfy the $so(4)$ commutation relations and $[\hat{r},\hat{S}_{ab}]\,=\,0$,

(ii) $\hat{U}_{\alpha \beta}$ and $\hat{U}^\dagger_{\alpha \beta}$ transforms as $so(4)$ bi-spinors,

(iii) $f(\hat{r})\,\hat{U}_{\alpha \beta}\,=\,\hat{U}_{\alpha \beta}\,f(\hat{r}+\lambda)$ and $f(\hat{r})\,\hat{U}^\dagger_{\alpha \beta}\,=\,\hat{U}_{\alpha \beta}\,f(\hat{r}-\lambda)$,

(iv) $\hat{U}_{\alpha \beta}$ and $\hat{U}^\dagger_{\alpha \beta}$ satisfy ordering relation (\ref{UU*order}).
\vskip0.5cm
%
Commutation relations in question follow directly from those among $\hat{U}$'s and $\hat{U}^\dagger$'s, as is shown in the appendix A2.These equations explicitly separate the commutators into Hermitian and anti-Hermitian parts. They lead to the following velocity commutators:
\begin{equation} \label{VV6} 
[\hat{V}_i,\,\hat{V}_j]\ =\ [\tilde{V}_{i},\,\tilde{V}_{j}] \ =\ -\frac{i}{\hat{r}^2}\,\hat{S}_{ij}\ +\ \frac{i\lambda }{2\hat{r}}\,\left( \{ \tilde{V}_i ,\hat{V}_j\} - \{\tilde{V}_j, \hat{V}_i\} \right) \,,
\end{equation} 
\begin{equation}\label{VV7}
[\hat{V}_k,\,\hat{V}_4]\ =\ - [\tilde{V}_k,\,\tilde{V}_4] \ =\ -\frac{i}{\hat{r}^2}\, \hat{S}_{k4}\ +\ \frac{i\lambda }{2\hat{r}} \left( \{\tilde{V}_k, \hat{V}_4\} + \{\hat{V}_k, \tilde{V}_4\}\right)   \,,
\end{equation}
\begin{equation} \label{VV8} 
[\hat{V}_i,\,\tilde{V}_j]\ =\ - [\tilde{V}_i,\,\hat{V}_j] \ =\ - \frac{i}{\lambda \hat{r}}\,\delta_{ij} \ +\ \frac{i\lambda }{2\hat{r}}\,\left(\{ \tilde{V}_i ,\tilde{V}_j\} + \{ \hat{V}_i ,\hat{V}_j\} \right) \,, 
\end{equation} 
\begin{equation} \label{VV9} 
[\hat{V}_k,\,\tilde{V}_4]\ =\ [\tilde{V}_k,\,\hat{V}_4] \ =\  -\frac{i\lambda }{2\hat{r}} \left( \{\tilde{V}_k,\tilde{V}_4\} - \{ \hat{V}_k,\,\hat{V}_4 \} \right) \,,
\end{equation}
\begin{equation} \label{VV10}
[\hat{V}_4,\,\tilde{V}_4]\ =\  \frac{i}{\lambda \hat{r}}\ -\ \frac{i\lambda }{\hat{r}}\,\left( \hat{V}^2_4 + \tilde{V}^2_4 \right)\,.
\end{equation}

Regarding monopoles is the most interesting the equation \eqref{VV6}. As is derived in the appendix A3, it reproduces exactly the result of \cite{mm}:
\begin{equation} \label{comVV}
[\hat{V}_i,\,\hat{V}_j] \ =\ -\frac{i\lambda(\kappa/2)}{\hat{r}(\hat{r}^2-\lambda^2)}\,\varepsilon_{ijk}\,\hat{S}_{k4} \,.
\end{equation}
Due to the obvious $SO(4)$ invariance, it can be extended to
\begin{equation} 
[\hat{V}_a,\,\hat{V}_b] \ =\ -\frac{i\lambda (\kappa/2)}{\hat{r}(\hat{r}^2-\lambda^2)}\,\varepsilon_{abcd}\,\hat{S}_{cd}\ \equiv\ i\,\hat{F}_{ab} \, ,
\end{equation}
that covers both equations (\ref{VV1}) and (\ref{VV2}).

The antisymmetric tensor $\hat{F}_{ab}$ describes the magnetic monopole field. Similarly, we could derive the formula for $[\hat{V}_a,\,\tilde{V}_b]\,=\,i\,\hat{G}_{ab}$ which is symmetric in $a,b$, and covers equations (\ref{VV3}-\ref{VV5}). The symmetric tensor $\hat{G}_{ab}$ is given in terms of anti-commutators $\{ \tilde{V}_a ,\tilde{V}_b\}$, $\{ \hat{V}_a, \tilde{V}_b \}$ and $\{ \hat{V}_a ,\hat{V}_b\}$. It can be decomposed into scalar part proportional to $\delta_{ab}$ and into trace-less tensor part. Such expression does not possesses a straightforward commutative limit and we leave any details or interpretations for the future.

We will now check whether the algebra of velocity operators is associative. In \cite{Szabo:2017yxd} it has been discussed that a theory formulated in a space uniformly filled with a magnetic monopole is nonassociative. On the other hand, we are dealing with an ordinary algebra of operators on a Hilbert space so it should be associative. Note that $ \varepsilon_{ijk} [ \hat{S}_{0i}, [ \hat{S}_{0j}, \hat{S}_{0k}]] = 0 $, but the velocity operators also contains the factor $r^{-1}$, which changes the overall commutators and makes their vanishing highly non-trivial. We need to evaluate the following commutator:
\begin{eqnarray} \label{VVV}
\varepsilon_{ijk} [ \hat{V}_i, [ \hat{V}_j , \hat{V}_k ]] &=&  \frac{\lambda \kappa}{2} \left[ \hat{V}_i, \frac{-i}{\hat{r} \left(\hat{r}^2 - \lambda^2 \right) } \hat{S}_{i4} \right] \\ \nonumber
&=& -i \frac{\lambda \kappa}{2} \left(\frac{1}{\hat{r}\left(\hat{r}^2 - \lambda^2 \right)} \left[ \hat{V}_i, \hat{S}_{i4} \right] + \left[\hat{V}_i , \frac{1}{\hat{r}\left(\hat{r}^2 - \lambda^2 \right)} \right] \hat{S}_{i4} \right)  \\ \nonumber
&=& -\frac{3  \lambda \kappa}{2 \hat{r} \left( \hat{r}^2 - \lambda^2 \right)} \hat{V}_4 + \frac{3  \lambda \kappa}{2 \hat{r} \left( \hat{r}^2 - \lambda^2 \right)} \hat{V}_4 = 0 .
\end{eqnarray}
In the first line we have utilized the form of $[\hat{V}_i, \hat{V}_j]$ from \eqref{comVV}. The commutator splits into two terms, the first of which is just a trivial relation following from \eqref{S}. The second term is more involving, details of the calculation are done in the appendix A4. The proof of the vanishing associator is now complete. 

\section{Conclusions}  \label{sec5}

We have investigated structure of the algebra of operators on the Hilbert space $\mathcal{H}_\kappa$ describing magnetic monopoles states of charge $\mu = - \frac{\kappa}{2}$. We have considered operators quadratic in left/right multiplication with auxiliary bosonic operators which have been used to defined both the underlying NC space $\textbf{R}_\lambda^3$ and the Hilbert space $\mathcal{H} = \sum \limits_{\kappa \in \mathbf{Z}} \oplus \mathcal{H}_{\kappa}$. There are 16 linearly independent operators, 15 of which has been realized as a $su(2,2)$ representation $\hat{S}_{AB}$, the last one being the center element identifying the corresponding subspace $\mathcal{H}_\kappa$. As it turned out, nearly all of this operators have been used in the previous studies of the model (each of them defined \textit{ad hoc}).

As long as the norm in $\mathcal{H}$ contains a weight function $f^{-1}(\hat{r})$, the elements of $\hat{S}_{AB}$ corresponding to boosts are not Hermitian with respect to it and have to be modified with a factor $\propto f(\hat{r})$. This, however, spoils the $su(2,2)$ structure, as we have discussed such modification introduces an infinite tower of factors $\delta^n_{\pm} f(\hat{r})$. We have also shown that the choice $f(\hat{r}) \propto \hat{r}^{-1}$, which is required by the norm \eqref{norm}, plays a special role –  it simplifies the resulting structure considerably. 

One of the consequences of this choice is that the (NC-deformed) Heisenberg algebra corresponds to that of a system containing a magnetic monopole of an arbitrary charge allowed by the Dirac quantization condition. Also, as we have shown in \eqref{VVV}, the associator vanishes and the system remains associative (and geometric \cite{mmna1, mmna2}). 

Surprisingly, the velocity operator $\hat{V}_a$ comes with a dual $\tilde{V}_a$. Their linear combinations $\hat{U}, \hat{U}^\dagger$ form a q-deformed commutator algebra with an $\hat{r}$-dependent factor $\hat{Q}\,=\, \sqrt{\frac{\hat{r}-\lambda}{\hat{r}+\lambda}}$, which approaches unity when $\hat{r} \gg \lambda$, as is the commutativity being restored. This connects the theory with the well-research field of q-deformed algebras. 

Commutators of the velocity operators and their duals can be expressed using their anti-commutators, elements of $\hat{S}_{AB}$ and powers of $\hat{r}^{-1}$ (contrary to $\delta^n_{\pm} \hat{r}^{-1}$). It shall be noted that while the commutator of elements of $\hat{V}_a$ (or $\tilde{V}_a$) alone define an antisymmetric tensor $\hat{F}_{ab}$, the commutator $[\hat{V}_a, \tilde{V}_b]$ defines a symmetric tensor $\hat{G}_{ab}$. Whether it has a gravitational interpretation or not is a possible line of future research. Another interesting option is to investigate operators $\mathcal{H}_\kappa \rightarrow \mathcal{H}_{\kappa' \neq \kappa}$, which should create/annihilate monopole charge. 

\subsection*{Acknowledgment}
This research was partially supported by COST Action MP1405 (S.K. and P.P.), project VEGA 1/0985/16 (P.P.) and the Irish Research Council funding (S.K.).

\section{Appendices}  \label{sec6}
\subsection*{A1:} \label{a1}
The velocity operator and its dual can be expressed using the operators defined in \eqref{U} as
\begin{eqnarray}
\hat{V}_k\ =\ \frac{i}{2} \left( \sigma^k_{\alpha \beta }\hat{U}_{\alpha \beta}\,-\,\sigma^{k*}_{\alpha \beta }\hat{U}^\dagger_{\alpha \beta} \right) \,,\ \ \ \hat{V}_4\ =\ \frac{1}{2} \left( \hat{U}_{\alpha\alpha }\,+\,\hat{U}^\dagger_{\alpha\alpha} \right)\,, \\ \nn
\tilde{V}_k\ =\ \frac{1}{2} \left( \sigma^k_{\alpha \beta }\hat{U}_{\alpha \beta}\,+\, \sigma^{k*}_{\alpha \beta }\hat{U}^\dagger_{\alpha \beta} \right)\,,\ \ \ \tilde{V}_4\ =\ \frac{i}{2} \left(  \hat{U}_{\alpha\alpha }\,-\, \hat{U}^\dagger_{\alpha\alpha } \right)\,.
\end{eqnarray}
The operators $\hat{U}_{\alpha \beta}$ and $\hat{U}^\dagger_{\alpha \beta}$ possess simple commutation relations
\begin{equation}\label{UU}
[\hat{U}_{\alpha \beta}, \hat{U}_{\gamma \delta} ]\ =\ [\hat{U}^\dagger_{\alpha \beta}, \hat{U}^\dagger_{\gamma \delta} ]\ =\ 0 \, ,
\end{equation}
\begin{equation}\label{UU*}
[\hat{U}_{\alpha \beta}, \hat{U}^\dagger_{\gamma \delta} ]\ =\ - \frac{1}{\hat{r}^2}\,\left( \hat{a}^+_\alpha \hat{a}_\gamma \,\delta_{\beta \delta} \,-\, \hat{b}^+_\delta  \hat{b}_\beta \delta_{\gamma \alpha} \right)\ +\ \frac{\lambda}{\hat{r}}\, \{ \hat{U}_{\alpha \beta}, \hat{U}^\dagger_{\gamma \delta}\} \,,
\end{equation}
Using the relations
\begin{equation} \label{UF}
f(\hat{r})\,\hat{U}_{\alpha \beta}\ =\ \hat{U}_{\alpha \beta}\,f(\hat{r}+\lambda )\,,\ \ \ f(\hat{r})\,\hat{U}^\dagger_{\alpha \beta}\ =\ \hat{U}^\dagger_{\alpha \beta}\,f(\hat{r}-\lambda)\,,
\end{equation}
is the proof of the first line in (\ref{UU}) simple
\begin{equation} \nn
\frac{1}{\hat{r}}\hat{a}^+_\alpha \hat{b}_\beta\,\frac{1}{\hat{r}} \hat{a}_\gamma^+ \hat{b}_\delta \ -\ 
\frac{1}{\hat{r}} \hat{a}_\gamma^+ \hat{b}_\delta \frac{1}{\hat{r}}\,\frac{1}{\hat{r}}\hat{a}^+_\alpha \hat{b}_\beta \ =\ \frac{1}{\hat{r}(\hat{r}-\lambda )}\,[\hat{a}^+_\alpha \hat{b}_\beta\,, \hat{a}^+_\gamma \hat{b}_\delta] \ =\ 0 \, .
\end{equation}
To prove the commutator (\ref{UU*}) is more involved. We separate the proof into a few steps:
\begin{equation}
[\hat{U}_{\alpha \beta}, \hat{U}^\dagger_{\gamma \delta} ]\ =\ \frac{1}{\hat{r}}\hat{a}^+_\alpha \hat{b}_\beta\,\frac{1}{\hat{r}} \hat{b}_\delta^+ \hat{a}_\gamma \ -\ \frac{1}{\hat{r}}  \hat{b}_\delta^+  \hat{a}_\gamma\,\frac{1}{\hat{r}}\hat{a}^+_\alpha \hat{b}_\beta  \nn 
\end{equation}
\begin{equation}
\ =\ \frac{1}{\hat{r}(\hat{r}-\lambda)}\hat{a}^+_\alpha \hat{b}_\beta\, \hat{b}_\delta^+ \hat{a}_\gamma\ -\ 
\frac{1}{\hat{r}(\hat{r}+\lambda )} \hat{a}_\gamma^+ \hat{b}_\delta \,\frac{1}{\hat{r}} \hat{b}_\delta^+ \hat{a}_\gamma \nn
\end{equation}
\begin{equation}
\ =\ \frac{1}{\hat{r}(\hat{r}^2-\lambda^2)} \left( (\hat{r}+\lambda)\,\hat{a}^+_\alpha \hat{b}_\beta\, \hat{b}_\delta^+ \hat{a}_\gamma\ -\ (\hat{r}-\lambda) \hat{b}_\delta^+  \hat{a}_\gamma\,\hat{a}^+_\alpha \hat{b}_\beta\right) \nn
\end{equation}
\begin{equation}\label{UU*0}
\ =\ \frac{1}{\hat{r}^2-\lambda^2}\,[\hat{a}^+_\alpha \hat{b}_\beta ,\,\hat{b}_\delta^+\hat{a}_\gamma ]\ +\ \frac{\lambda }{\hat{r}(\hat{r}^2-\lambda^2)}\,\{ \hat{a}^+_\alpha \hat{b}_\beta ,\,\hat{b}_\delta^+\hat{a}_\gamma \} \,.
\end{equation}
The first term in (\ref{UU*0}) can be reduced using the relation
\begin{equation}\label{UU*1}
[\hat{a}^+_\alpha \hat{b}_\beta ,\,\hat{b}_\delta^+\hat{a}_\gamma ]\ =\ - \hat{a}^+_\alpha \hat{a}_\gamma \,\delta_{\beta \delta} \ -\ \hat{b}_\delta^+\hat{b}_\beta \,\delta_{\gamma \alpha} \,,   
\end{equation}
whereas the second term in (\ref{UU*0}) can be rewritten as follows
\begin{equation}\label{UU*2}
\frac{\lambda }{\hat{r}(\hat{r}^2-\lambda^2)}\,\left( \hat{a}^+_\alpha \hat{b}_\beta \,\hat{b}_\delta^+\hat{a}_\gamma \ +\ \hat{b}_\delta^+\hat{a}_\gamma \hat{a}^+_\alpha \hat{b}_\beta \right) =\ \frac{\lambda }{\hat{r}^2-\lambda^2} \left( (\hat{r}-\lambda)\,\hat{U}_{\alpha \beta }\, \hat{U}^\dagger_{\gamma \delta} \ +\ (\hat{r}+\lambda)\,\hat{U}^\dagger_{\gamma \delta }\,\hat{U}_{\alpha \beta } \right) \, .
\end{equation}
Using (\ref{UU*1}) and (\ref{UU*2}), the relation for the commutator (\ref{UU*}) can be rewritten as
\begin{equation} 
 [\hat{U}_{\alpha \beta}, \hat{U}^\dagger_{\gamma \delta} ]\ =\ - \frac{1}{\hat{r}^2-\lambda^2} \left( \hat{a}^+_\alpha \hat{a}_\gamma \,\delta_{\beta \delta} \ +\ \hat{b}_\delta^+\hat{b}_\beta \,\delta_{\gamma \alpha} \right)\ -\ \frac{\lambda^2 }{\hat{r}^2-\lambda^2}\,[\hat{U}_{\alpha \beta}, \hat{U}^\dagger_{\gamma \delta} ] \nn
\end{equation}
%
\begin{equation} \label{UU*3}
 +\ \frac{\lambda \hat{r} }{\hat{r}^2-\lambda^2}\,\{ \hat{U}_{\alpha \beta}, \hat{U}^\dagger_{\gamma \delta}\} \,. 
\end{equation}
On both sides of \eqref{UU*3} there appears the same commutator $[\hat{U}, \hat{U}^\dagger ]$. Extracting it and dividing the equation by $1+\frac{\lambda^2}{r^2 - \lambda^2} (= \frac{r^2}{r^2-\lambda^2})$ we obtain a remarkable relation
\begin{equation} \label{UU*4}
 [\hat{U}_{\alpha \beta}, \hat{U}^\dagger_{\gamma \delta} ]\ =\ -\frac{1}{\hat{r}^2} \left( \hat{a}^+_\alpha \hat{a}_\gamma \,\delta_{\beta \delta} \ +\ \hat{b}_\delta^+\hat{b}_\beta \,\delta_{\gamma \alpha} \right) +\ \frac{\lambda}{\hat{r}}\,\{ \hat{U}_{\alpha \beta}, \hat{U}^\dagger_{\gamma \delta}\} \,, 
\end{equation}

\subsection*{A2:}\label{a2}
Commuatation relation for the velocity and its dual can be expressed using the commutators of \eqref{U}:
\begin{equation} \label{VV1} 
[\hat{V}_i,\,\hat{V}_j]\ =\ [\tilde{V}_i,\,\tilde{V}_j] \ =\  \frac{1}{4} \left( \sigma^i_{\alpha\beta} \sigma^{j*}_{\gamma \delta}\,[\hat{U}_{\alpha \beta}, \hat{U}^\dagger_{\gamma \delta} ]\ -\ h.c. \right)\,, 
\end{equation} 
\begin{equation}\label{VV2}
[\hat{V}_k,\,\hat{V}_4]\ =\ - [\tilde{V}_k,\,\tilde{V}_4] \ =\  \frac{i}{4} \left( \sigma^k_{\alpha\beta}\, [\hat{U}_{\alpha \beta}, \hat{U}^\dagger_{\gamma\gamma} ]\ +\ h.c. \right) \,,
\end{equation}
\begin{equation} \label{VV3} 
[\hat{V}_i,\,\tilde{V}_j]\ =\ - [\tilde{V}_i,\,\hat{V}_j] \ =\  \frac{i}{4} \left( \sigma^i_{\alpha\beta} \sigma^{j*}_{\gamma \delta}\,[\hat{U}_{\alpha \beta}, \hat{U}^\dagger_{\gamma \delta} ]\ +\ h.c. \right)\,, 
\end{equation} 
\begin{equation} \label{VV4} 
[\hat{V}_k,\,\tilde{V}_4]\ =\ [\tilde{V}_k,\,\hat{V}_4] \ =\  \frac{1}{4} \left( \sigma^k_{\alpha\beta}\, [\hat{U}_{\alpha \beta}, \hat{U}^\dagger_{\gamma\gamma} ]\ -\ h.c. \right) \,,
\end{equation}
\begin{equation} \label{VV5}
[\hat{V}_4,\,\tilde{V}_4]\ =\ -\frac{i}{2}\, [\hat{U}_{\alpha\alpha},\hat{U}^\dagger_{\gamma\gamma}]\,.
\end{equation}
Since the Pauli matrices are Hermitian we can rewrite the commutator term in (\ref{VV1}) and (\ref{VV3}) as follows 
\begin{eqnarray} \nn
&&\sigma^i_{\alpha\beta} \sigma^j_{\delta\gamma}\,[\hat{U}_{\alpha \beta}, \hat{U}^\dagger_{\gamma \delta} ]\ =\ \,\sigma^i_{\alpha\beta} \sigma^j_{\delta\gamma} \left( -\frac{1}{\hat{r}^2}(\hat{a}^+_\alpha \hat{a}_\gamma \,\delta_{\beta\delta} \,+\,\hat{b}_\delta^+\hat{b}_\beta \,\delta_{\gamma \alpha})\,+\, \frac{\lambda}{\hat{r}}\,\{ \hat{U}_{\alpha \beta}, \hat{U}^\dagger_{\gamma \delta} \} \right)  \\ \nn
 & &=\ - \frac{1}{\hat{r}^2}\,(\hat{a}^+_\alpha \hat{a}_\alpha\,+\, \hat{b}^+_\gamma \hat{b}_\gamma )\,\delta_{ij} \ -\ \frac{2i}{\hat{r}^2}\,\hat{S}_{ij}\ +\ \frac{\lambda }{\hat{r}}\,\{ \tilde{V}_i - i \hat{V}_i,\, \tilde{V}_j + i \hat{V}_j\} \,,    \\ \label{comUU*1}
 &&=\ - \frac{2}{\lambda \hat{r}}\,\delta_{ij} \ +\ \frac{\lambda }{\hat{r}}\,\left(\{ \tilde{V}_i ,\tilde{V}_j\} + \{ \hat{V}_i ,\hat{V}_j\} \right)-\ \frac{2i}{\hat{r}^2}\,\hat{S}_{ij}\ +\ \frac{i\lambda }{\hat{r}}\,\left( \{ \tilde{V}_i ,\hat{V}_j\}- \{\tilde{V}_j, \hat{V}_i\} \right).
 \end{eqnarray}
Here the first two terms in the last equation are symmetric and the other two are antisymmetric in $i,j$.
Similarly: 
\begin{eqnarray}  
\sigma^k_{\alpha\beta}\, \left[\hat{U}_{\alpha \beta}, \hat{U}^\dagger_{\gamma\gamma} \right]\ &= &\ - \frac{1}{\hat{r}^2} \sigma^k_{\alpha\beta} \left( \hat{a}^+_\alpha \hat{a}_\gamma \,\delta_{\beta\gamma} \,+\,\hat{b}_\gamma^+\hat{b}_\beta \,\delta_{\gamma \alpha} \right) +\ \frac{\lambda}{\hat{r}}\,\sigma^k_{\alpha\beta}\,\{ \hat{U}_{\alpha \beta}, \hat{U}^\dagger_{\gamma\gamma} \} \\ \nn
&=&-\ \frac{2}{\hat{r}^2}\, \hat{S}_{k4}\ +\ \frac{\lambda }{\hat{r}} \left( \{\tilde{V}_k, \hat{V}_4\} + \{\hat{V}_k, \tilde{V}_4\}\right) + \frac{i\lambda }{\hat{r}} \left( \{\tilde{V}_k,\tilde{V}_4\} - \{ \hat{V}_k,\,\hat{V}_4 \} \right),  \\ \nn
\left[ \hat{U}_{\alpha\alpha }, \hat{U}^\dagger_{\gamma\gamma} \right] &=& - \frac{1}{\hat{r}^2}\, \left(\hat{a}^+_\alpha \hat{a}_\alpha\,+\, \hat{b}^+_\gamma \hat{b}_\gamma \right)\ +\ \frac{\lambda}{\hat{r}}\, \{ \hat{U}_{\alpha \beta}, \hat{U}^\dagger_{\gamma\gamma} \}  \\ 
&=&\ - \frac{2}{\lambda \hat{r}}\ +\ \frac{2\lambda }{\hat{r}}\left( \hat{V}^2_4\,+\,\tilde{V}^2_4 \right) \,.   
\end{eqnarray}
\subsection*{A3:}\label{a3}
We want to evaluate $[\hat{V}_i,\,\hat{V}_j]$ which is antisymmetric in $(i,j)$. Therefore, it is equivalent to work with the expression
\begin{eqnarray} \nn 
\varepsilon_{ijk}\,[\hat{V}_i,\,\hat{V}_j]\ &=&\ \frac{1}{4} \varepsilon_{ijk} \sigma^i_{\alpha\beta} \sigma^{j}_{\delta \gamma} \left( \,[\hat{U}_{\alpha \beta}, \hat{U}^\dagger_{\delta \gamma} ]\ -[\hat{U}_{\gamma \delta}, \hat{U}^\dagger_{\beta \alpha} ] \right) \\ \nn
&=&\ \frac{i}{2}\,\sigma^k_{\alpha \beta} \left( [\hat{U}_{\alpha\delta}, \hat{U}^\dagger_{\beta \delta} ]\,-\,[\hat{U}_{\delta\beta}, \hat{U}^\dagger_{\delta\alpha} ] \right) ,
\end{eqnarray}
where we used the identity $\varepsilon_{ijk} \sigma^i_{\alpha\beta} \sigma^{j}_{\delta \gamma}\,=\, i(\sigma^k_{\alpha\delta} \delta_{\gamma \beta} - \sigma^k_{\gamma \beta}\delta_{\alpha\delta})$.
Explicitly, the commutators are given as
\begin{eqnarray}
\sigma^k_{\alpha \beta} [\hat{U}_{\alpha \delta}, U^\dagger_{\beta \delta} ] &=& \sigma^k_{\alpha \beta} \left( \frac{1}{\hat{r}} \hat{a}_\alpha^+ \hat{b}_\delta \frac{1}{\hat{r}} \hat{a}_\beta \hat{b}^+_\delta - \frac{1}{\hat{r}}\hat{a}_\beta \hat{b}^+_\delta \frac{1}{\hat{r}} \hat{a}^+_\alpha \hat{b}_\delta \right) \\ \nonumber
&=& \frac{\sigma^k_{\alpha \beta} \hat{a}^+_\alpha \hat{a}_\beta}{\hat{r}\left(\hat{r}^2 -\lambda^2 \right)} \left( (\hat{r}+\lambda) \hat{b}_\delta \hat{b}_\delta^+ - (\hat{r}-\lambda) \hat{b}^+ \hat{b} \right) \\ \nonumber
&=& - \frac{\lambda\sigma^k_{\alpha \beta}  \hat{a}^+_\alpha \hat{a}_\beta }{\hat{r}\left(\hat{r}^2 -\lambda^2 \right)} (\hat{C}+2), \\ \nonumber
\sigma^k_{\alpha \beta}  [ \hat{U}_{\delta \beta}, \hat{U}^\dagger_{\delta \alpha} ] &=&  \sigma^k_{\alpha \beta}  \left( \frac{1}{\hat{r}}\hat{a}^+_\delta \hat{b}_\beta \frac{1}{\hat{r}} \hat{a}_\delta \hat{b}^+_\alpha - \frac{1}{\hat{r}}\hat{a}_\delta \hat{b}_\alpha^+ \frac{1}{\hat{r}} \hat{a}_\delta^+ \hat{b}_\beta \right) \\ \nonumber
&=&  \frac{\sigma^k_{\alpha \beta} \hat{b}^+_\alpha \hat{b}_\beta}{\hat{r}\left(\hat{r}^2 -\lambda^2 \right)} \left( (\hat{r}+\lambda) \hat{a}^+_\delta \hat{a}_\delta - (\hat{r}-\lambda) \hat{a}_\delta \hat{a}^+_\delta \right) \\ \nonumber
&=&   \frac{\lambda\sigma^k_{\alpha \beta}  \hat{b}^+_\alpha \hat{b}_\beta}{\hat{r}\left(\hat{r}^2 -\lambda^2 \right)}  (\hat{C}+2) ,
\end{eqnarray}
where $\hat{C} = \hat{a}_\delta ^+ \hat{a}_\delta - \hat{b}_\delta ^+ \hat{b}_\delta$ is the representation of the central element introduced above. Combining both commutators we obtain
\begin{equation} \nn
\varepsilon_{ijk}\,[\hat{V}_i,\,\hat{V}_j]\ =\ -\frac{i\lambda (\hat{C} + 2)}{2\hat{r}(\hat{r}^2-\lambda^2)}\,\sigma^k_{\alpha\gamma}\,(\hat{a}_\alpha^+ \hat{a}_\gamma\,+\,\hat{b}_\alpha^+ \hat{b}_\gamma )\ =\ -\frac{i\lambda (\hat{C} + 2)}{\hat{r}(\hat{r}^2-\lambda^2)}\,\hat{S}_{k4} \,, 
\end{equation}
The final step is to realize that $\hat{C}+2 = \kappa$ on the considered subspace $\mathcal{H}_\kappa$. 
\subsection*{A4:} \label{a4}
\begin{eqnarray} \nn
\frac{\lambda \kappa}{2}  \left[\hat{V}_i , \frac{1}{\hat{r}\left(\hat{r}^2 - \lambda^2 \right)}\right] \hat{S}_{i4} &=& \frac{i \lambda \kappa}{4 \hat{r}}\left(  \left[ \hat{a}^+_\alpha b_\beta, \frac{1}{\hat{r}\left(\hat{r}^2 - \lambda^2 \right)}\right] -  \left[ \hat{a}_\beta b_\alpha^+ , \frac{1}{\hat{r}\left(\hat{r}^2 - \lambda^2 \right)}\right] \right) \sigma^i_{\alpha \beta} \hat{S}_{i4} \\ \nonumber
&=& \frac{i \lambda \kappa }{4 \hat{r}^2} \Big( \left( \frac{1}{(\hat{r} - \lambda)(\hat{r}-2\lambda)} - \frac{1}{(\hat{r}-\lambda) (\hat{r}+\lambda)} \right)\hat{a}^+_\alpha \hat{b}_\beta \\ \nonumber
&&-\left( \frac{1}{(\hat{r}+2\lambda) (\hat{r}+\lambda)} - \frac{1}{(\hat{r}-\lambda) (\hat{r}+\lambda)}\right)\hat{a}_\beta \hat{b}^+_\beta \Big) \sigma^i_{\alpha \beta} \hat{S}_{i4} \\ \nonumber
&=&\frac{3i \lambda^2   \kappa  }{4 \hat{r}^2 (\hat{r}^2 - \lambda^2 )} \left( \frac{\hat{a}^+_\alpha \hat{a}_\alpha + \hat{b}^+_\alpha \hat{b}_\alpha -4 }{\hat{r}-2\lambda} \hat{a}^+_\alpha \hat{b}_\alpha + \frac{\hat{a}^+_\alpha \hat{a}_\alpha + \hat{b}^+_\alpha \hat{b}_\alpha +4 }{\hat{r}+2\lambda} \hat{a}_\alpha \hat{b}^+_\alpha \right) \\ 
&=& \frac{3i \lambda  \kappa}{2 \hat{r} (\hat{r}^2 - \lambda^2)} \frac{\hat{S}_{04} }{2\hat{r}}=  \frac{3i \lambda \kappa}{2 \hat{r} (\hat{r}^2 - \lambda^2)} \hat{V}_4
\end{eqnarray}
We have used relations \eqref{UF} to obtain the second line and the scalar Fierz identity for the Pauli matrices to obtain the second to the last line.

\end{document}